\documentclass[useAMS,usenatbib]{mn2e}
\usepackage[dvips]{graphicx}
\usepackage{epsfig}
\usepackage{rotating}
\usepackage{natbib}
\usepackage{color}
\usepackage{mathrsfs}
\usepackage{amssymb}
\usepackage[caption=false]{subfig}
\usepackage{eqnarray,amsmath}
\usepackage{hyperref}

\newcommand{\Msun}{\, {\rm M_{\odot}}}

\title[Origin of red-sequence galaxies]{The origin of the red sequence galaxy population in the EAGLE simulation}
\author[C.A.~Correa, J.~Schaye and J.W.~Trayford]
{Camila A.~Correa$^1$\thanks{E-mail: correa@strw.leidenuniv.nl}, Joop~Schaye$^1$ and James W. Trayford$^1$\\
 $^1$ Leiden Observatory, Leiden University, P.O. Box 9513, 2300 RA Leiden, The Netherlands}

\pagerange{\pageref{firstpage}--\pageref{lastpage}} \pubyear{2013}

\def\LaTeX{L\kern-.36em\raise.3ex\hbox{a}\kern-.15em
    T\kern-.1667em\lower.7ex\hbox{E}\kern-.125emX}

\begin{document}
\maketitle

\begin{abstract}
We investigate the evolution in colour and morphology of the progenitors of red-sequence galaxies in the EAGLE cosmological hydrodynamical simulation. We quantify colours with $u^{*}-r^{*}$ intrinsic magnitudes and morphologies with a measure of the stellar kinematics. The time when galaxies moved onto the red sequence depends on their morphology. Disc-type galaxies tend to have become red during the last 3 Gyr, while elliptical-type galaxies joined the red sequence earlier, with half the sample already being red 5 Gyr ago. The time-scale, $\tau_{\rm{Green}}$, of colour transition through the `green valley' depends weakly on the galaxy's morphological type. Elliptical-type galaxies cross the green valley slightly faster ($\tau_{\rm{Green}}{\approx}1$ Gyr) than disc-type galaxies ($\tau_{\rm{Green}}{\approx}$1.5 Gyr). While $\tau_{\rm{Green}}$ is similar for central and satellite galaxies, for satellites $\tau_{\rm{Green}}$ decreases with increasing stellar mass to host-halo mass ratio. Coupled with our finding that galaxies tend to become green after becoming satellites, this indicates that satellite-specific processes are important for quenching red-sequence galaxies. The last time central, elliptical-type red-sequence galaxies left the blue cloud is strongly correlated with the time the luminosity of the central black hole peaked, but this is not the case for discs. This suggests that AGN feedback is important for quenching ellipticals, particularly centrals, but not for discs. We find only a weak connection between transformations in colour and morphology.
\end{abstract}


\section{Introduction}

Most galaxies in the local Universe can be divided into two distinct populations, a spiral or disc-dominated population and a spheroidal or bulge-dominated population (e.g. \citealt{Willett13}). Large-scale surveys have shown that these two populations also differ in colour, star formation, gas content and surrounding environment (e.g \citealt{Baldry04,Schawinski07,Brammer09}). Spiral galaxies are mostly actively star forming with blue colours and preferentially inhabit low-density regions, whereas spheroidal or elliptical-type galaxies are passive or weakly star-forming with red colours and are commonly found in high-density regions (e.g. \citealt{Dressler1980,Strateva01}). 

These two populations of galaxies are broadly referred to as the `blue cloud' and the `red sequence', respectively, in the optical colour-magnitude diagram. Between the red sequence (hereafter RS) and the blue cloud, there is a sparsely populated region with a smaller number density of galaxies. This region is known as the `green valley' (e.g. \citealt{Martin07,Wyder07}). 

While red galaxies have older stellar populations than blue galaxies (e.g. \citealt{Kauffmann03}), the stellar populations in green-valley galaxies are older than those in blue-cloud galaxies and younger than those in RS galaxies (\citealt{Pan13}). This supports the evolutionary scenario where blue galaxies evolve into red ones, transitioning through the green valley (e.g. \citealt{Martin07,Gonzalves12,Schawinski14}), and is also reflected in the luminosity density of blue galaxies, which has significantly decreased since $z{\approx}1$, while the luminosity density of red galaxies has remained roughly constant (\citealt{Bell04,Faber07}).

It is difficult to establish the main mechanism that shifts galaxies from the blue cloud to the red sequence. Recent results suggest the existence of two contrasting evolutionary pathways through the green valley, where elliptical-type galaxies likely underwent rapid quenching accompanied by a morphological transformation from disk to spheroid, and passive disc-type galaxies likely quenched by a slow exhaustion of the gas over several Gyr, driven by environmental processes (e.g. \citealt{Schawinski14,Smethurst15}). 

For the pathway of elliptical-type galaxies, the classical theoretical scenario indicates that when disc-type galaxies of similar mass merge, the remnant has an elliptical morphology (\citealt{Toomre1977}). Mergers may trigger a starburst with associated black hole growth and explosive AGN feedback that reddens and quenches the galaxy (e.g. \citealt{Hernquist1989,DiMatteo05}). However, several observational (e.g. \citealt{Grogin05,Gabor09,Cisternas11,Schawinski12,Rosario15}) and theoretical (\citealt{McAlpine17,Steinborn18}) studies have suggested that mergers may not play a dominant role in triggering AGN, and thus cannot fully explain the formation of red galaxies. Many other studies have proposed AGN feedback to be the primary mechanism that causes both the suppression or quenching of star formation in massive galaxies, and gives rise to the correlation of the central black hole mass with the galactic bulge mass (e.g. \citealt{Martin07,Schawinski07b,McConnell13}). Mergers that change the galaxy morphology may, however, still be able to quench the galaxy without the need for AGN feedback, by forming a stellar spheroid that stabilizes any remaining or newly formed disc against fragmentation into bound star-forming clumps (e.g. \citealt{Martig09}).

The pathway of disc-type galaxies not only suggests that the transition to being passive is not necessarily accompanied by a change in morphology (as shown by recent surveys, e.g. \citealt{Moran06,Bamford09,Wolf09,vanderWel10,Masters10}) but it is also driven by environmental processes, that affect disc-type central galaxies differently from satellites. Central galaxies sit at the minimum of the host halo's potential well, where the gas densities and cooling rates tend to be highest, while satellite galaxies orbit the central galaxy. The most studied environmental mechanisms that are considered to contribute to satellite quenching are ram pressure stripping (e.g. \citealt{Gunn1972,Abadi99,Bahe13}), starvation (also called strangulation, which occurs when the gaseous halo is removed, thus cutting off the star formation gas supply; \citealt{Larson80}, see also \citealt{Balogh00}) and harassment (high speed galaxy `fly-by' gravitational interactions, e.g. \citealt{Moore96,Moore99}). These mechanisms have been found to predominantly affect satellite galaxies residing in cluster-size haloes (e.g. \citealt{Bahe15}). 

In this work we aim to shed light on these matters by analysing how morphology can elucidate the different evolutionary paths of the progenitors of RS galaxies in the EAGLE simulation (\citealt{Schaye15,Crain15}). In a recent study (\citealt{Correa17a}), we analysed the scatter plot of optical intrinsic $u^{*}-r^{*}$ galaxy colour versus stellar mass for different morphologies, centrals and satellites in the EAGLE simulation. The scatter plot not only revealed the well known red sequence and blue cloud (whose excellent match to observations was initially shown by \citealt{Trayford15}), it also revealed that the blue cloud and red sequence comprise mostly disc- and elliptical-type galaxies, respectively. At intermediate masses, the red sequence consists, however, of a more morphologically diverse population of satellite and central galaxies.

In this work we use the morphological parametrization of \citet{Correa17a} and the intrinsic $u^{*}-r^{*}$ colours of \citet{Trayford15} to show that the way in which EAGLE galaxies move onto the red sequence depends on their morphological history. This work expands upon previous analyses of morphology and star formation history of EAGLE galaxies. \citet{Trayford16} showed that the red sequence of EAGLE galaxies starts to build-up around $z{=}1$, due to the quenching of low-mass satellite galaxies at the faint end, and of more massive central galaxies by their active galactic nuclei (AGN) at the bright end. The works of \citet{Clauwens17} and \citet{Trayford18} have investigated the origin of the Hubble sequence by analysing how kinematically defined disc and spheroidal structures form and evolve in EAGLE. \citet{Trayford18} showed that star formation occurs predominantly in disc structures throughout most of cosmic time, however morphological transformations remove stars from discs at a specific rate of $\sim 0.07$ Gyr$^{-1}$ at $z<5$, thus establishing the low-redshift morphological mix. \citet{Clauwens17} identified three phases of galaxy formation, with low-mass galaxies growing in a disorganised way, intermediate-mass galaxies evolving towards a disc-dominated morphology driven by in-situ star formation, and high-mass galaxies forming a spheroidal component that grows mostly by accretion of stars.

While this work was being refereed, we became aware of the study of Wright et al. (in prep.), which uses the EAGLE simulation to determine the time over which galaxies transition the green valley in the colour-mass diagram, or leave the main sequence of star formation. The work of Wright et al. is largely complementary to this study, they correlate the quenching timescale with the galaxies' gas fractions, AGN activity and host halo mass. They use, however, a lower time resolution than this work, and do not separate galaxies according to morphology.

The layout of the paper is as follows. In Section 2 we describe the EAGLE simulation used in this study, particularly the aspects related to galaxy colour and morphology parametrization. Section 3 defines the red sequence galaxy population in EAGLE. Section 4 analyses the past evolution of colour and morphology of RS galaxies. We calculate green valley transition time scales in Section 4.2 and investigate the impact of environment, AGN feedback and morphological transformation on the past colour evolution of the progenitors of red sequence galaxies in Section 5. Finally, our findings are summarised in Section 6.

\section{Methodology}
\subsection{The EAGLE simulation}

Throughout this work we analyse the cosmological, hydrodynamical simulation RefL100N1504 (hereafter Ref), from the EAGLE simulation series. RefL100N1504 represents a cosmological volume of 100 comoving Mpc on a side that was run with a modified version of GADGET 3 (\citealt{Springel05}), an $N$-Body Tree-PM smoothed particle hydrodynamics (SPH) code, which was modified to use an updated formulation of SPH, new time stepping and new subgrid physics (see \citealt{Schaye15}, for a complete description). RefL100N1504 contains 1504$^{3}$ dark matter (as well as gas) particles, with initial gas and dark matter particle masses of $m_{\rm{g}}=1.8\times 10^{6}\Msun$ and $m_{\rm{dm}}=9.7\times 10^{6}\Msun$, respectively, and a Plummer-equivalent gravitational softening length of $\epsilon=0.7$ proper kpc at $z=0$. It assumes a $\Lambda$CDM cosmology derived from the {\it{Planck-1}} data (\citealt{Planck}), $\Omega_{\rm{m}}=1-\Omega_{\Lambda}=0.307$, $\Omega_{\rm{b}}=0.04825$, $h=0.6777$, $\sigma_{8}=0.8288$, $n_{s}=0.9611$, and a primordial mass fraction of hydrogen of $X=0.752$.

Dark matter haloes and the self-bound substructures within them hosting individual galaxies are identified using the Friends-of-Friends and SUBFIND algorithms respectively (\citealt{Springel01,Dolag09}). In each halo, the `central' galaxy is the galaxy closest to the centre (minimum of the potential), usually the most massive. The remaining galaxies within the halo are its satellites. 

We link dark matter haloes through consecutive snapshots following the merger trees from the EAGLE public database (\citealt{McAlpine16}). These merger trees were created using the D-Trees algorithm of \citet{Jiang14}, see also \citet{Qu17}. In the tree, the main progenitor is defined as the progenitor with the largest `branch mass', i.e., the mass of its progenitors summed across all earlier outputs, as proposed by \citet{DeLucia07}. We determine a galaxy's assembly history by following its main progenitor through 145 output redshifts between $z=0$ and $z=4$. This high time resolution is achieved by using the 145 RefL100N1504 `snipshots', which contain only the main particle properties but are output with much higher frequency than the regular snapshots.

\subsection{Galaxy colour}

We use the galaxy colours computed by \citet{Trayford15}, who adopted the GALAXEV population synthesis model of \citet{Bruzual03} to determine the spectral energy distribution per unit initial stellar mass of a simple stellar population for a discrete grid of ages and metallicities. We use the intrinsic (i.e. rest-frame and dust-free) $u^{*}-r^{*}$ colours (with the $*$ referring to intrinsic), where the blue cloud and red sequence in the colour-stellar mass relation appear clearly separated. \citet{Trayford15} showed that these colours are in agreement with observational data. To determine the galaxy colours (as well as stellar mass) we follow \citet{Schaye15} and use spherical apertures of 30 proper kpc.

\subsection{Galaxy morphology}

\begin{figure} 
\centering
\includegraphics[angle=0,width=0.48\textwidth]{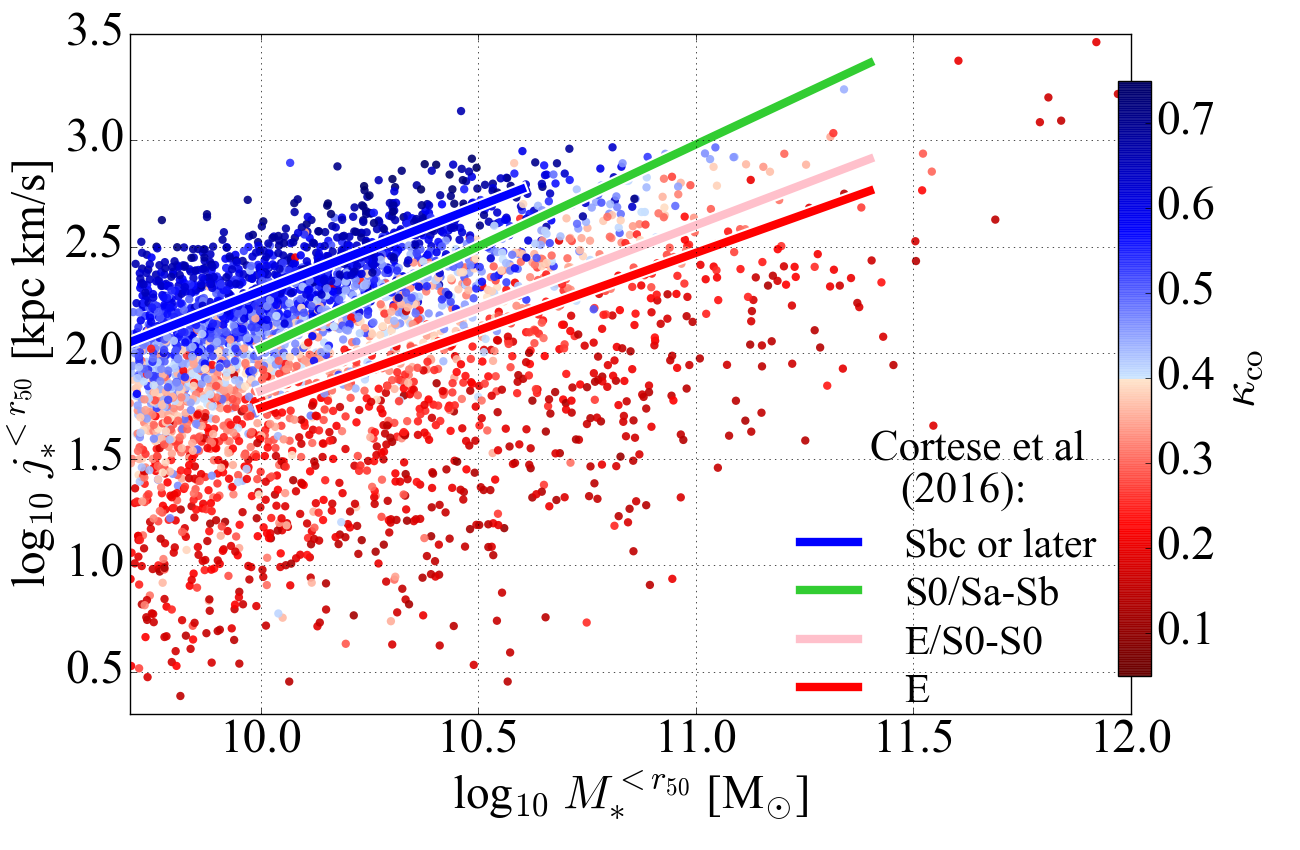}
\caption{Comparison between the stellar mass-specific angular momentum ($M_{*}^{<r_{50}}-j_{*}^{<r_{50}}$) of EAGLE galaxies (dots) and the median $M_{*}^{<r_{50}}-j_{*}^{<r_{50}}$ relations of Cortese et al. (2016, solid lines). Each dot in the figure is an individual galaxy coloured according to its corotation parameter, $\kappa_{\rm{co}}$, following the colour bar on the right. The solid lines correspond to the $M_{*}^{<r_{50}}-j_{*}^{<r_{50}}$ median relation of Sbc or later-type (blue line), S0/Sa-Sb-type (green line), E/S0-S0-type (pink line) and E-type (red line) galaxies as calculated in Cortese et al. (2016). At fixed stellar mass, as $j_{*}^{<r_{50}}$ decreases, galaxies change morphology from disc (blue dots with $\kappa_{\rm{co}}\geq 0.4$) to elliptical (red dots with $\kappa_{\rm{co}}<0.4$). The $M_{*}^{<r_{50}}-j_{*}^{<r_{50}}$-visual morphology relation of \citet{Cortese16} follows the same behavior, indicating that $\kappa_{\rm{co}}$ is able to trace visual morphology.}
\label{comparison}
\end{figure}

To quantify galaxy morphology, we use the stellar kinematics-based morphological indicator computed by \citet{Correa17a}. In \citet{Correa17a} we defined $\kappa_{\rm{co}}$ as the fraction of kinetic energy invested in ordered {\it{corotation}}, 

$$\kappa_{\rm{co}}=\frac{K_{\rm{corot}}}{K}=\frac{1}{K}\sum_{i}^{r<30\rm{kpc}}\frac{1}{2}\,m_i\left[L^{>0}_{z,i}/(m_i\,R_i)\right]^{2},$$

\noindent where $m_{i}$ is the mass of stellar particle $i$, $K(=\sum_{i}^{r<30\rm{kpc}}\frac{1}{2}m_iv_i^2)$ is the total kinetic energy, $L^{>0}_{z,i}$ is the particle angular momentum along the direction of the total angular momentum of the stellar component of the galaxy ($\vec{L}$) and $R_{i}$ is the projected distance to the axis of rotation ($\vec{L}$). In the equation, the sum is over all star particles that follow the direction of rotation of the galaxy (i.e. $L^{>0}_{z,i}$ positive) and are located within a spherical radius of 30 pkpc centered on the minimum of the potential. \citet{Correa17a} showed that high-$\kappa_{\rm{co}}$ galaxies ($\kappa_{\rm{co}}\ge 0.4$) tend to be rotation-dominated, disc-shaped galaxies, whereas low-$\kappa_{\rm{co}}$ galaxies ($\kappa_{\rm{co}}< 0.4$) tend to be dispersion-dominated and spherical. Therefore $\kappa_{\rm{co}}=0.4$ was used to separate galaxies that look disky from those that look elliptical. Hereafter we refer to disc-type galaxies with $\kappa_{\rm{co}}\ge 0.4$ as disc galaxies and elliptical-type galaxies with $\kappa_{\rm{co}}<0.4$ as elliptical galaxies. 

Although it has been shown that stellar kinematics provide a physically motivated morphological classification (e.g. \citealt{Fall1983,Kormendy1993,Kormendy04,Romanowsky12,Snyder15,Teklu15}), it has also been concluded that kinematic morphological indicators sometimes fail to discriminate between different types of objects (\citealt{Emsellem07,Emsellem11}). 

Recently, \citet{Cortese16} used a sample of galaxies from the SAMI survey to show that the scatter in the stellar mass-specific angular momentum relation is strongly correlated with optical morphology. They performed a visual morphological classification using SDSS DR9 colour images following the scheme of \citet{Kelvin14}, and referred to early-type galaxies with a bulge as ellipticals (`E'), early-types with discs as `S0s', late-types with a disc component as `Sc or later' and late-types with disc plus bulge as `Sa-Sb'. They calculated the median stellar mass-specific angular momentum relation for each galaxy type and showed that visual morphology drives the spread in the relation (also reported in \citealt{Romanowsky12} and \citealt{Obreschkow14}). 

We use the stellar mass, specific angular momentum and visual morphology relation of \citet{Cortese16} to test whether $\kappa_{\rm{co}}$ correlates with the visual morphology relation of SAMI galaxies. We calculate the light-weighted specific angular momentum, $j_{*}^{<r_{50}}$, and stellar mass, $M_{*}^{<r_{50}}$, within the galaxy half-mass radius $r_{50}$ (following \citealt{Lagos17}, who showed that by doing so $j_{*}^{<r_{50}}$ can be directly compared to SAMI data). $j_{*}^{<r_{50}}$ is calculated as follows

$$j_{*}^{<r_{50}}=\frac{\sum_{i}^{<r_{50}}L_{{\rm{r}},i}({\bf{r}}_{i}-{\bf{r}}_{\rm{CoM}})\times({\bf{v}}_{i}-{\bf{v}}_{\rm{CoM}})}{\sum_{i}^{<r_{50}}F_{{\rm{r}},i}},$$

\noindent where ${\bf{r}}_{i}$, ${\bf{v}}_{i}$ and ${\bf{r}}_{\rm{CoM}}$, ${\bf{v}}_{\rm{CoM}}$ are the position and velocity vectors of particle $i$ and the centre of mass, respectively, and $L_{{\rm{r}},i}$ is the r-band total luminosity of particle $i$. Note that in this section $M_{*}^{<r_{50}}$ refers to the total mass of star particles within the galaxy half-mass radius, however in the rest of the work stellar masses denoted by $M_{*}$ correspond to the total mass of star particles within 30 pkpc. 

Fig.~\ref{comparison} shows the $M_{*}^{<r_{50}}-j_{*}^{<r_{50}}$ relation. The dots show individual galaxies from the Ref model, coloured according to their corotation parameter (following the colour bar on the right). For comparison, the solid lines show the median $M_{*}^{<r_{50}}-j_{*}^{<r_{50}}$ relations from \citet{Cortese16}, and correspond to Sbc or later-type (blue line), S0/Sa-Sb-type (green line), E/S0-S0-type (pink line) and E-type (red line) galaxies. At fixed stellar mass it can be seen that as $j_{*}^{<r_{50}}$ decreases, galaxies change morphology from disc (blue dots with $\kappa_{\rm{co}}\geq 0.4$) to elliptical (red dots with $\kappa_{\rm{co}}<0.4$). The $M_{*}^{<r_{50}}-j_{*}^{<r_{50}}$-visual morphology relation of \citet{Cortese16} follows the same behavior, indicating that $\kappa_{\rm{co}}$ is able to trace visual morphology.

Fig.~\ref{comparison} also shows that at fixed stellar mass, pure elliptical EAGLE galaxies tend to have lower $j_{*}^{<r_{50}}$ than the median values of \citet{Cortese16}. However, \citet{Cortese16} point out that the inclination correction likely overestimates the projection effects in the SAMI galaxies. The $M_{*}^{<r_{50}}-j_{*}^{<r_{50}}$ red solid line should therefore be taken as an upper limit.

\begin{figure*} 
\includegraphics[angle=0,width=0.75\textwidth]{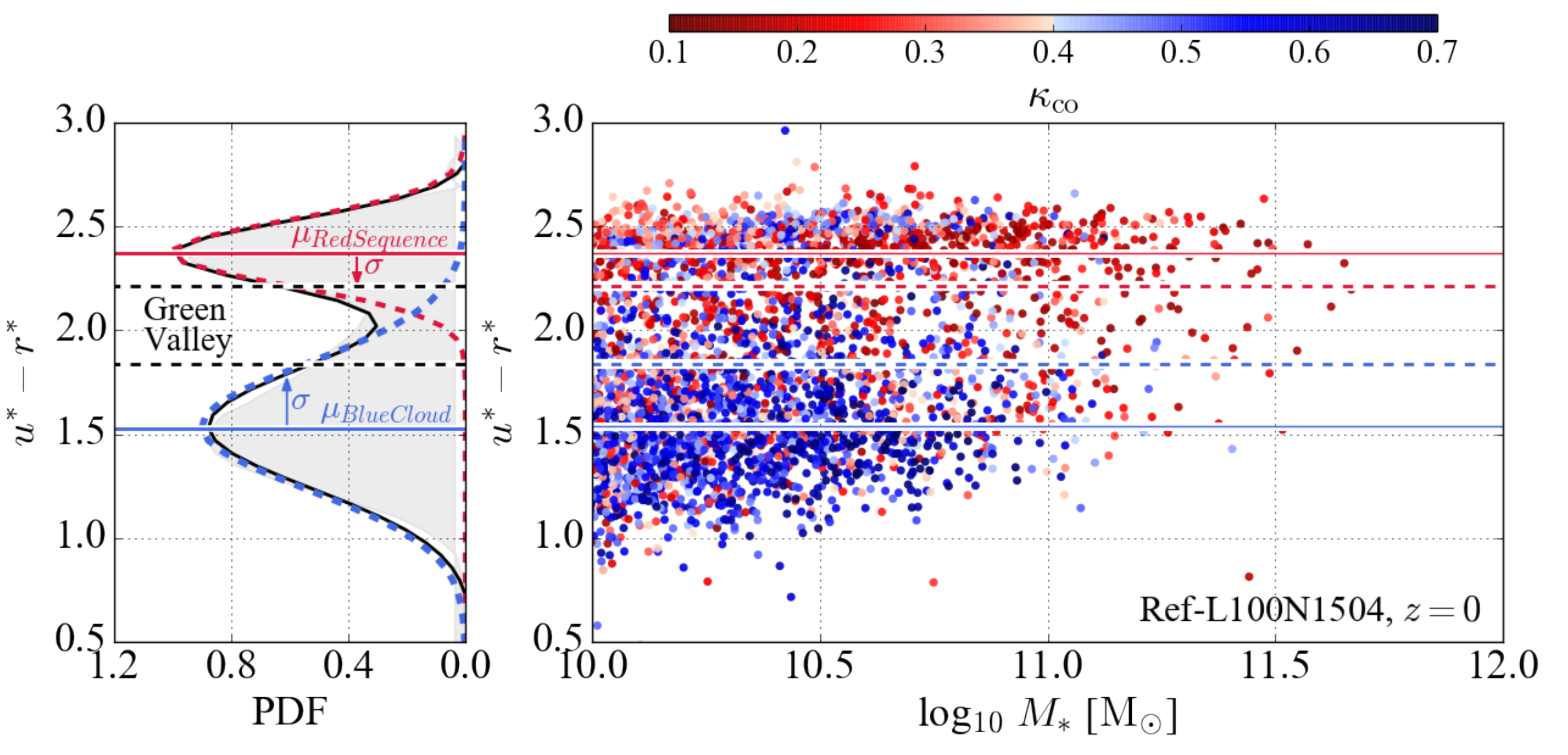}
\vspace{-0.2cm}
\caption{Left panel: probability density distribution of the $u^{*}-r^{*}$ colour (grey shaded region) for $z=0$ EAGLE galaxies with stellar masses larger than $10^{10}\Msun$. Red and blue dashed lines show the best-fit double gaussian function, from which we determine the means and standard deviations of the red sequence and blue cloud galaxy populations used to define the green valley. Right panel: $u^{*}-r^{*}$ colour-mass diagram of $z=0$ EAGLE galaxies. Each dot in the panel is an individual galaxy coloured by $\kappa_{\rm{co}}$ according to the colour bar on the top. The mean colour values, $\mu_{\rm{red}}$ and $\mu_{\rm{blue}}$, of the red and blue populations are shown as red and blue solid lines, respectively, and the boundaries given by eqs.~(\ref{red_seq_def}-\ref{blue_cloud_def}) are shown as dashed lines. }
\label{fig_images}
\end{figure*}

\begin{figure} 
\begin{center}
\includegraphics[angle=0,width=0.44\textwidth]{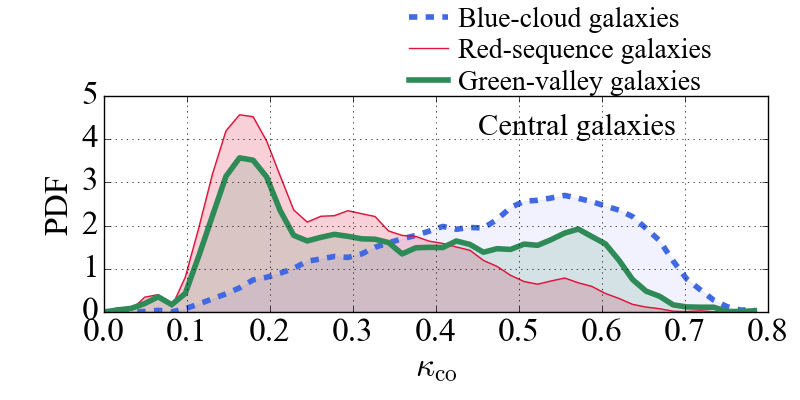}\\
\vspace{-1.4cm}
\includegraphics[angle=0,width=0.44\textwidth]{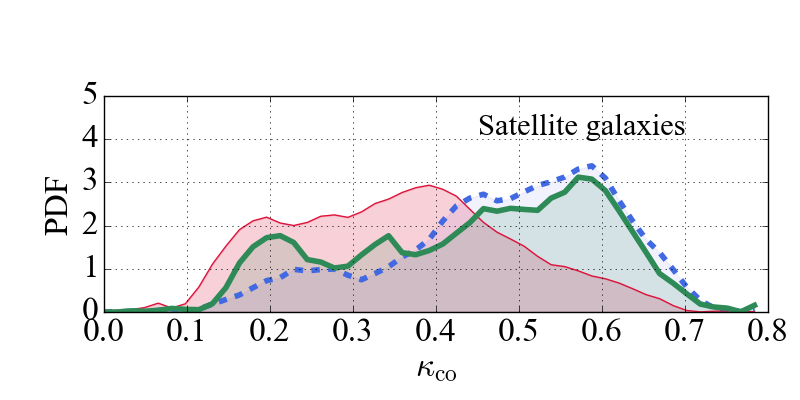}\\
\vspace{-0.2cm}
\caption{Probability distribution functions of $\kappa_{\rm{co}}$ for central (top panel) and satellite (bottom panel) galaxies. In each panel galaxies are separated into red-sequence, green-valley and blue-cloud according to their $u^{*}-r^{*}$ colours. Blue-cloud galaxies tend to have much greater rotational support than red-sequence and green-valley galaxies, but for the case of satellites, red and green galaxies are morphologically diverse.}
\label{kappa_fraction}
\end{center}
\end{figure}

\section{The galaxy samples}

\subsection{The colour-mass diagram}

In this work we explore the evolution and properties of RS galaxies more massive than $10^{10}\Msun$. \citet{Schaye15} and \citet{Furlong15} showed that resolution effects cause an upturn in the passive fraction at lower masses. The right panel of Fig.~\ref{fig_images} shows the $u^{*}-r^{*}$ colour-mass diagram, where individual galaxies are coloured by $\kappa_{\rm{co}}$ to show that morphology correlates with both colour and mass (see \citealt{Correa17a} for a discussion). It can be seen that while galaxies with $u^{*}-r^{*}{\geq}2$ are mostly ellipticals, there is also a significant fraction of disc galaxies for $M_{*}<10^{11}\Msun$. We note that contrary to the distribution of colours, the distribution of $\kappa_{\rm{co}}$ in EAGLE is not bimodal.

The left panel shows the probability distribution function (PDF) of colour (grey shaded region). We fit a double gaussian to the distribution and obtain the means ($\mu_{\rm{red}}, \mu_{\rm{blue}})=(2.37,1.52$) and standard deviations ($\sigma_{\rm{red}}, \sigma_{\rm{blue}})=(0.16,0.29$) of the red-sequence and blue-cloud populations shown using red and blue dashed lines, respectively. We next define two boundaries in colour, $(u^{*}-r^{*})_{\rm{red}}$ and $(u^{*}-r^{*})_{\rm{blue}}$, as 1$\sigma$ away from the population mean colour, $\mu$, as follows

\begin{eqnarray}\label{red_seq_def1}
(u^{*}-r^{*})_{\rm{red}} &\equiv& \mu_{\rm{red}}(z)-\sigma_{\rm{red}}(z),\\\label{blue_cloud_def1}
(u^{*}-r^{*})_{\rm{blue}} &\equiv& \mu_{\rm{blue}}(z)+\sigma_{\rm{blue}}(z).
\end{eqnarray}

\noindent A galaxy with $u^{*}-r^{*}>(u^{*}-r^{*})_{\rm{red}}$ is classified as red, and a galaxy with $u^{*}-r^{*}<(u^{*}-r^{*})_{\rm{blue}}$ is classified as blue. 

We analyse the redshift evolution in the range $0\le z\le 4$ of the double gaussian best-fitting parameters $\mu_{\rm{red}}$, $\mu_{\rm{blue}}$, $\sigma_{\rm{red}}$, $\sigma_{\rm{blue}}$ and find the redshift dependence to be $\propto z^{0.6}$, in agreement with \citet{Trayford16}. We also analysed the stellar mass dependence of the colour distribution. For RS galaxies there is a slight dependence of $u^{*}-r^{*}$ on $M_{*}$, driven by the galaxies' metallicity (\citealt{Trayford16}), but the slope is flatter ($u^{*}-r^{*}{\propto}0.07\log_{10}(M_{*}/\Msun)$) than the one obtained by \citet{Schawinski14} (${\propto}0.25\log_{10}(M_{*}/\Msun)$) using SDSS+GALEX+Galaxy Zoo data. For simplicity, we thus set the slope to zero for both the RS and blue cloud, yielding colour boundaries of

\begin{eqnarray}\label{red_seq_def}
(u^{*}-r^{*})_{\rm{red}} = 2.21-0.65z^{0.6},\\\label{blue_cloud_def}
(u^{*}-r^{*})_{\rm{blue}} = 1.81-0.59z^{0.6}.
\end{eqnarray}

The right panel of Fig.~\ref{fig_images} shows the means as a function of stellar mass of the red and blue populations ($\mu_{\rm{red}}(M_{*},z{=}0)$, $\mu_{\rm{blue}}(M_{*},z{=}0)$) as red and blue solid lines, respectively, and the boundaries given by eqs.~(\ref{red_seq_def}-\ref{blue_cloud_def}) as dashed lines. The green valley is defined as the space in the colour-mass diagram between the blue and red population boundaries. The red sequence galaxy population comprises 1178 galaxies, which corresponds to $33\%$ of the total EAGLE galaxy population with stellar masses larger than $10^{10}\Msun$. While most of this red sequence population consists of elliptical galaxies ($68\%$, out of which $49\%$ are centrals and $51\%$ satellites), the remaining disc galaxies are mostly satellites ($70\%$ against $30\%$ centrals). The green valley contains 631 galaxies ($18\%$ of the total sample) and the blue cloud 1753 galaxies ($49\%$ of the total).

\subsection{Distribution of galaxy morphology}

In Section 2.3 we showed that in the specific angular momentum-stellar mass plane the corotation parameter, $\kappa_{\rm{co}}$, follows the visual morphology classification of \citet{Cortese16}, with low$-\kappa_{\rm{co}}$ galaxies being S0-E type and high$-\kappa_{\rm{co}}$ galaxies being Sbc type. In this section we analyse how the distribution of $\kappa_{\rm{co}}$ changes when galaxies are separated using the colour boundaries defined in Section 3.1.
 
Fig.~\ref{kappa_fraction} shows the PDF of $\kappa_{\rm{co}}$ for RS, blue-cloud and green-valley galaxies separated into centrals (top panel) and satellites (bottom panel). The $\kappa_{\rm{co}}$ PDF of RS, blue-cloud and green-valley galaxies are shown using thin red, dashed blue and thick green solid lines, respectively. In the top panel it can be seen that there is a peak at $\kappa_{\rm{co}}=0.17$ in the distribution of central RS and green-valley galaxies, indicating that most of these galaxies have dispersion-dominated kinematics and predominantly a spherical morphology. The distribution of disc central galaxies does not exhibit a prominent peak, instead it increases smoothly towards large $\kappa_{\rm{co}}$ values. The bottom panel of Fig.~\ref{kappa_fraction} shows that the distribution of satellite galaxies differs from that of centrals. Most red satellite galaxies are characterized by $\kappa_{\rm{co}}$ values around 0.4 thus exhibiting neither a prominent disc nor bulge. The $\kappa_{\rm{co}}$ PDF of green satellite galaxies is bimodal, with peaks at $\kappa_{\rm{co}}=0.19$ and $0.58$, indicating that this sample is morphologically diverse.

\section{Origin of the red sequence galaxy population}

In this section we analyse the evolution in colour and morphology of RS galaxies between $z=0$ and 4, and show that the way galaxies populate the present-day red sequence depends on their morphological history.


\begin{figure} 
\hspace{-0.3cm}
\includegraphics[angle=0,width=0.25\textwidth]{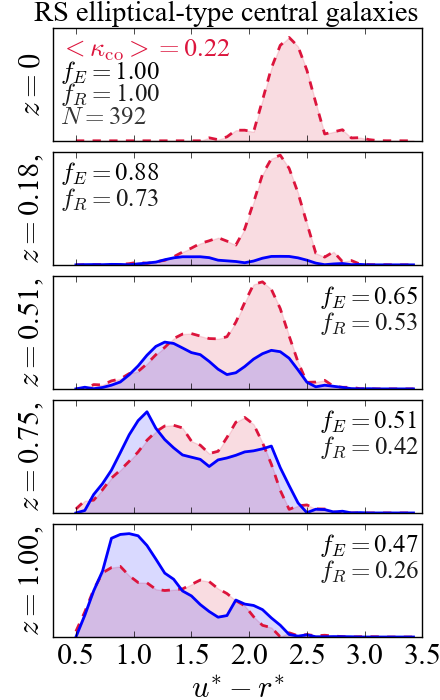}
\hspace{-0.5cm}
\includegraphics[angle=0,width=0.25\textwidth]{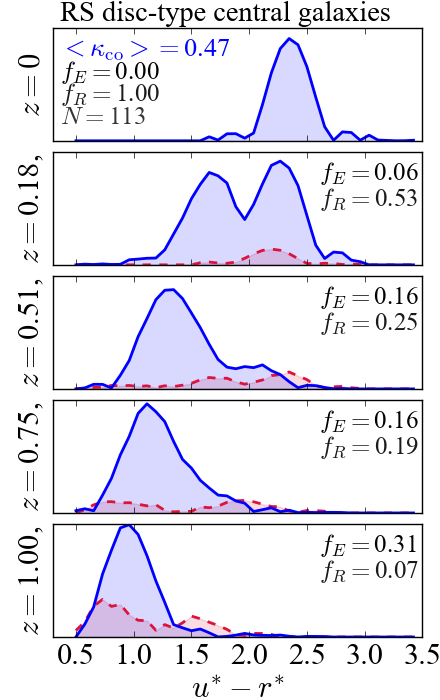}\\
\hspace*{-0.3cm}
\includegraphics[angle=0,width=0.25\textwidth]{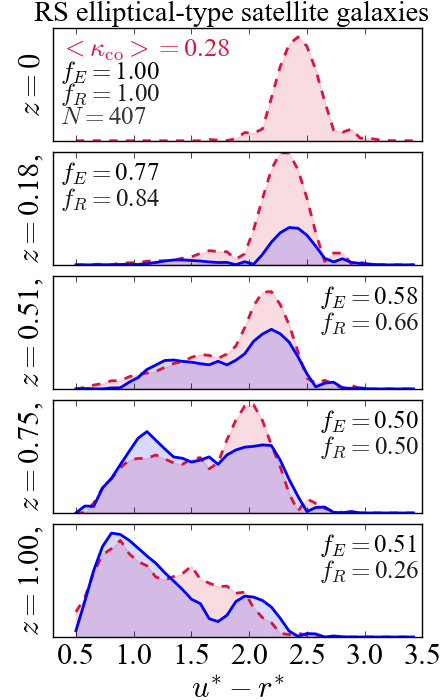}
\hspace{-0.5cm}
\includegraphics[angle=0,width=0.25\textwidth]{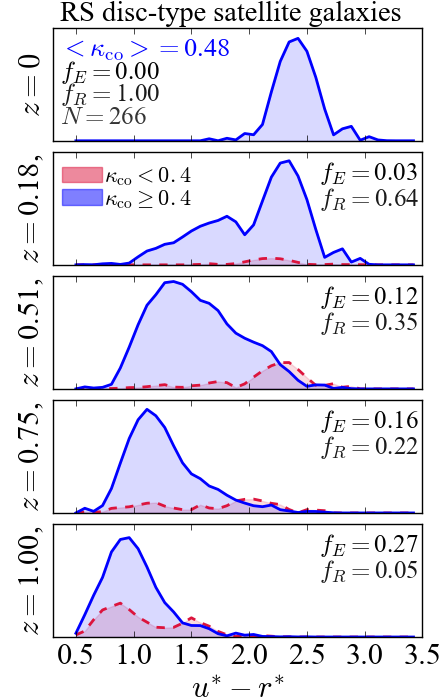}
\caption{$u^{*}-r^{*}$ colour distribution of the progenitors of $z=0$ red-sequence elliptical central and satellites galaxies (top-left and bottom-left panels) and $z=0$ red-sequence disc central and satellites galaxies (top-right and bottom-right panels). In each panel the curves show the $u^{*}-r^{*}$ distributions separating disc (blue histograms) from elliptical (red-histograms) galaxies to highlight the morphological evolution of the galaxies' progenitors. The legends indicate the fraction of progenitors that are elliptical ($f_{\rm{E}}; \kappa_{\rm{co}}<0.4$) and that are in the red sequence ($f_{\rm{R}}$; using the red sequence definition given by eq. 3). The progenitors of red-sequence disc galaxies were mostly disky and for $z{\gtrsim}0.3$ mostly blue. Among the progenitors of red-sequence elliptical, the fraction that are red and the fraction that are elliptical are much larger than for the progenitors of disc red-sequence galaxies.} 
\label{morpho_distribution}
\end{figure}

\begin{figure}
\centering 
\includegraphics[angle=0,width=0.42\textwidth]{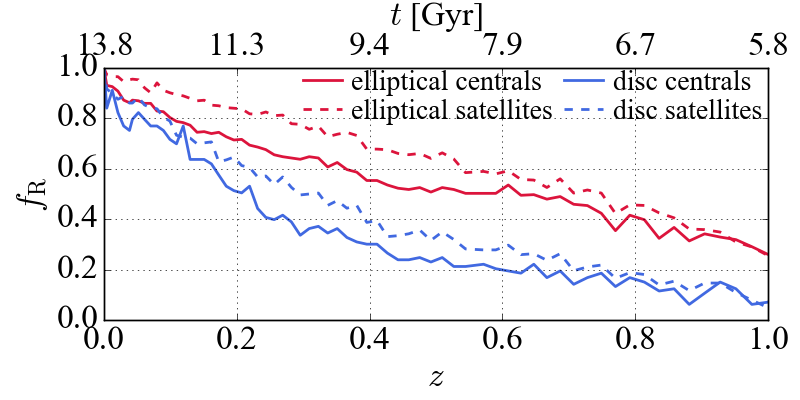}\\
\vspace*{-0.55cm}
\includegraphics[angle=0,width=0.42\textwidth]{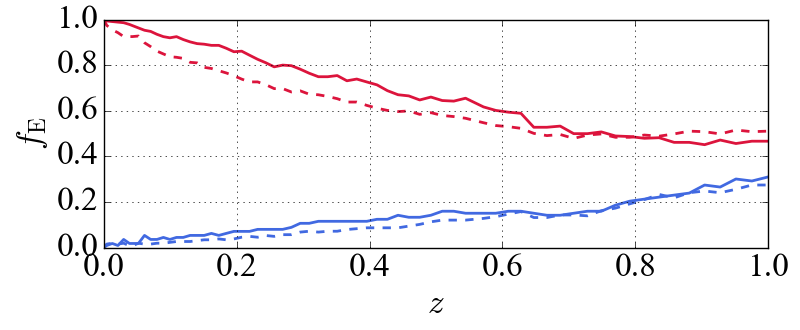}
\vspace{-0.4cm}
\caption{Evolution of the fraction of red-sequence progenitors that are on the red sequence ($f_{\rm{R}}$, top panel; using the red sequence definition given by eq. 3) or that are elliptical ($f_{\rm{E}}; \kappa_{\rm{co}}<0.4$, bottom panel) for $z=0$ disc (blue lines) and elliptical (red lines) galaxies separated into centrals (solid lines) and satellites (dashed lines). The top panel shows that the progenitors of elliptical satellite galaxies quenched slightly earlier than elliptical centrals. The panel also shows that the progenitors of most disc galaxies quench much later. The bottom panel shows that half the same of $z=0$ red-sequence elliptical galaxies were discs at $z=0.7$, and a significant fraction of red-sequence disc galaxies were ellipticals at $z=1$.}
\label{fE_fR_evolution}
\end{figure}

\subsection{Evolution in colour and morphology}

The EAGLE RS galaxy population is composed of 1178 galaxies with stellar masses larger than $10^{10}\Msun$, out of which $32\%$ are disc and $68\%$ are elliptical, according to our kinematic classification. We follow the merger history of this galaxy sample, and store the $u^{*}-r^{*}$ colour and $\kappa_{\rm{co}}$ parameter of the progenitors at each output redshift. We separate the progenitors by their morphological type and calculate their $u^{*}-r^{*}$ colour distributions.

The panels of Fig.~\ref{morpho_distribution} show the $u^{*}-r^{*}$ colour distributions of elliptical RS central (i.e. $\kappa_{\rm{co}}{<}0.4$, top left panel) and satellite (bottom left panel) galaxies, and the colour distributions of disc RS central (i.e. $\kappa_{\rm{co}}{\geq}0.4$, top right panel) and satellite (bottom right panel) galaxies at $z{=}0$. The panels show, for each subcategory of RS galaxy at $z=0$, i.e. elliptical/disc and central/satellite, the colour distributions of their progenitors separated by their morphology, so that blue and red histograms correspond to disc and elliptical progenitor galaxies, respectively. The legends indicate the fraction of progenitors that are elliptical ($f_{\rm{E}}$) and that are red ($f_{\rm{R}}$), according to the red sequence definition of eq.~(\ref{red_seq_def}).

The colour evolution of the progenitors of RS disc central and satellite galaxies indicates that most were preferentially discs during the redshift range $z=0-1$, and formed part of the blue-cloud at $z{\gtrsim}0.3$. This means that in most disc RS galaxies, star formation was quenched relatively recently, during the last 3.3 Gyr ($z=0-0.28$). Conversely, elliptical central and satellite RS galaxies are more likely to have undergone a morphological transformation, with half of the galaxies' progenitors having been discs at $z{\gtrsim}0.75$. This can be seen from the increasing number of disc galaxies with redshift, shown as blue distributions. Compared to disc RS galaxies, most elliptical galaxies move onto the RS earlier, with ${\gtrsim}53\%$ (${\gtrsim}66\%$ for satellites) of the sample being red for $z{\leq}0.5$ (5.2 Gyr ago). 

Fig.~\ref{fE_fR_evolution} shows the evolution of the fraction of progenitors that are elliptical ($f_{\rm{E}}; \kappa_{\rm{co}}<0.4$, bottom panel) and that are in the red sequence ($f_{\rm{R}}$, top panel) for $z=0$ RS disc (blue lines) and elliptical (red lines) galaxies separated into centrals (solid lines) and satellites (dashed lines). The top panel shows clearly that the progenitors of elliptical satellite galaxies quench slightly earlier than elliptical centrals, with half the sample being red by $z=0.75$, while for the progenitors of elliptical centrals, half the sample is red by $z=0.5$. The panel also shows that the progenitors of most disc galaxies quench much later, with half the sample of disc central and satellites being red by $z=0.2$ and by $z=0.3$, respectively.

The bottom panel of Fig.~\ref{fE_fR_evolution} shows not only that half the sample of $z=0$ RS elliptical galaxies were discs ($\kappa_{\rm{co}}\geq 0.4$) at $z=0.7$, but also that a significant fraction of RS disc galaxies (${\sim}30\%$) were ellipticals at $z=1$.

\begin{figure} 
\includegraphics[angle=0,width=0.49\textwidth]{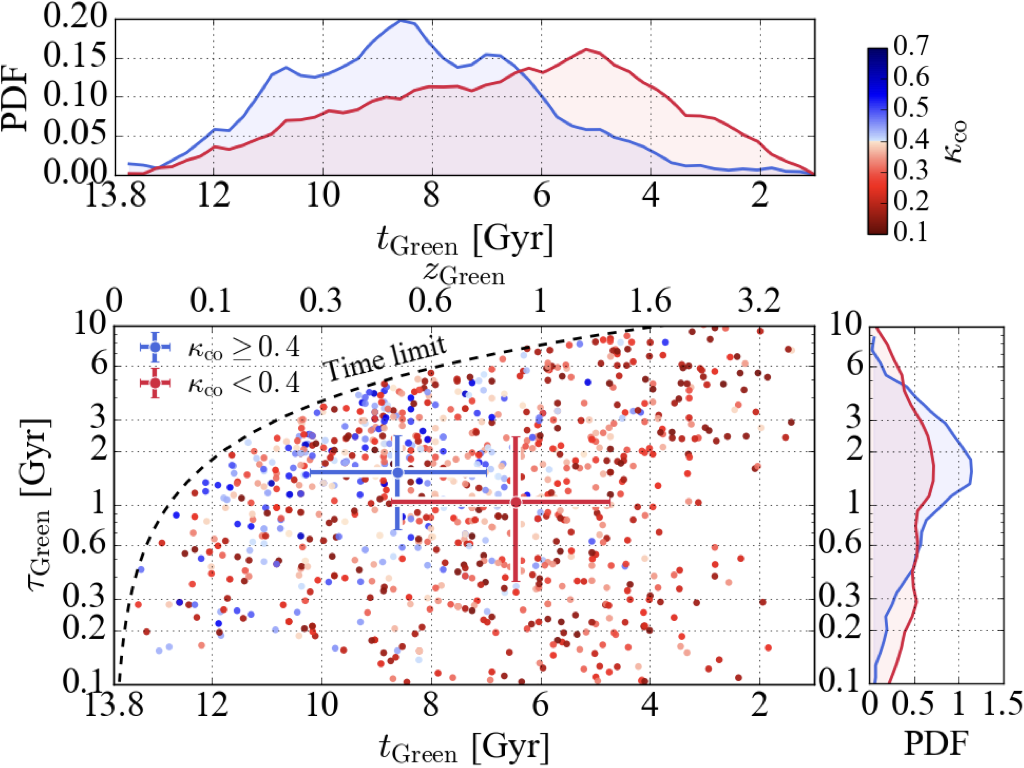}
\vspace{-0.3cm}
\caption{Bottom left panel: Green valley transitioning time-scale ($\tau_{\rm{Green}}$) for galaxies that form part of the red sequence at $z=0$, as a function of the {\it{last time}} they entered the green valley ($t_{\rm{Green}}$, shown as cosmic age along the bottom x-axis and redshift along the top x-axis). Each dot in the panel is a galaxy coloured according to its kinematic morphology at $z=0$ following the colour bar on the top right of the figure, with blue representing disc galaxies and red elliptical. The larger symbols with error bars correspond to the median values of $\tau_{\rm{Green}}$ and $t_{\rm{Green}}$, and the error bars correspond to the $25$th$-75$th percentiles. The dashed black line indicates the maximum possible value of $\tau_{\rm{Green}}$. Disc RS galaxies tend to have entered the green valley earlier than ellipticals. Top panel: probability density function (PDF) of $t_{\rm{Green}}$ for disc (blue distributions) and elliptical galaxies (red distributions). Bottom right panel: PDF of $\tau_{\rm{Green}}$ for disc (blue distributions) and elliptical galaxies (red distributions).}
\label{quench_time}
\end{figure}

\subsection{Green valley transition time-scale}

It has been suggested that the manner in which galaxies quench and cross the green valley depends on their morphological type (\citealt{Schawinski14}), with late-type (spiral) galaxies quenching slowly, and early-type (elliptical) galaxies quenching rapidly, transitioning the green valley in ${\sim}1$ Gyr (\citealt{Wong12}). In this section we follow the assembly history of elliptical and disc RS galaxies and estimate the time, $\tau_{\rm{Green}}$, these galaxies took to cross the green valley.

To obtain $\tau_{\rm{Green}}$ for each individual galaxy, we determine the time, $t_{\rm{Green}}$, when the galaxy had a $u^{*}-r^{*}$ colour redder than $(u^{*}-r^{*})_{\rm{blue}}$ (given by eq.~\ref{blue_cloud_def}), i.e. when it became green. The galaxy can return to the blue cloud multiple times before transitioning the green valley, so we define $t_{\rm{Green}}$ to be the cosmic time corresponding to the {\it{last time}} the galaxy was blue. Next, we determine the {\it{last time}} that the galaxy was green, i.e. it had a $u^{*}-r^{*}$ colour redder than $(u^{*}-r^{*})_{\rm{red}}$ (given by eq.~\ref{red_seq_def}). We refer to this time as $t_{\rm{Red}}$, and define $\tau_{\rm{Green}}=t_{\rm{Red}}-t_{\rm{Green}}$, with $t_{\rm{Red}}{>}t_{\rm{Green}}$.

The bottom left panel of Fig.~\ref{quench_time} shows $\tau_{\rm{Green}}$ as a function of $t_{\rm{Green}}$ (shown as cosmic age along the bottom x-axis and redshift along the top x-axis). Each dot in the panel is a galaxy coloured according to the colour bar on the top right of the figure, which indicates kinematic morphology at $z=0$. The symbols with error bars correspond to the median values of  $\tau_{\rm{Green}}$ and $t_{\rm{Green}}$, and the error bars correspond to the $25$th$-75$th percentiles. The top and bottom-right panels of the figure show the $t_{\rm{Green}}$ and $\tau_{\rm{Green}}$ PDFs for disc (blue distributions) and elliptical galaxies (red distributions). 

From the bottom-right panel it can be seen that although disc galaxies transitioned the green valley over a median time-scale of ${\approx}1.5$ Gyr, and most elliptical galaxies were quenched slightly faster, over a median timescale of ${\approx}$1 Gyr. However, the large spread of the distributions indicates that the correlation of $\tau_{\rm{Green}}$ with morphology is weak. Although the top panel also shows that the $t_{\rm{Green}}$ distributions also have a large spread, it can be seen that most disc galaxies transitioned the green valley during the last 5 Gyr, while most elliptical galaxies were quenched at earlier times.

Our median timescales for galaxies to cross the green valley are similar to those of \citet{Nelson17}, who reported $\tau_{\rm{green}}{\approx}1.6$ Gyr for galaxies from the IllustrisTNG simulations. \citet{Trayford16} obtained higher $\tau_{\rm{green}}$ values than this work. They also used the EAGLE simulation and found it to be approximately 2 Gyr for all galaxies, independent of mass. This difference is due to the method followed to calculate the $u^{*}-r^{*}$ colour boundaries (given by eqs.~\ref{red_seq_def}-\ref{blue_cloud_def}). We fit a double gaussian to the $u^{*}-r^{*}$ colour distribution at every output redshift (as in \citealt{Nelson17}) to determine the boundaries, whereas \citet{Trayford16} defined their $u^{*}-r^{*}$ colour cuts in a more ad-hoc way and obtained longer transition times. When implementing the $u^{*}-r^{*}$ colour cuts of \citet{Trayford16}, we reproduce their results.

In this section we conclude that the green valley transition timescale is weakly correlated with morphology. However, some observational studies have reached a different conclusion. \citet{Schawinski14} used the morphological classification from the Galaxy Zoo project (\citealt{Lintott08}), and showed that the colour distributions of green-valley elliptical galaxies can be modelled by a quenching time-scale, $\tau_{\rm{q}}$, of $\sim$100 Myr, whereas for disc galaxies a longer time-scale of ${\sim}2.5$ Gyr is required (see also \citealt{Smethurst15}). \citet{Nogueira18} calculated star formation quenching time-scales of $z{\sim}0.8$ galaxies from the COSMOS field and used a morphological classification based on principal component analysis of the galaxy structure. They found shorter quenching time-scales for the elliptical, irregular and merger morphologies or $\sim 100$ Myr compared to those of disc galaxies of $\sim 300$ Myr, which are shorter than those of \citet{Schawinski14}.

The discrepancy between the green valley transition timescale, $\tau_{\rm{green}}$, calculated in this work, and the quenching time-scale, $\tau_{\rm{q}}$, of \citet{Schawinski14} or \citet{Nogueira18} may signal a deficiency of the model, but it could also be due to the fact that different quantities are being measured. \citet{Schawinski14} (as well as \citealt{Nogueira18}) assumed a star formation rate (SFR) history of the form SFR$(t)\propto e^{-t/\tau_{\rm{q}}}$, with $\tau_{\rm{q}}$ the quenching time-scale, and convolved it with \citet{Bruzual03} population synthesis model spectra to obtain dust-corrected $NUV-u$ colour tracks. They found that the colours of their SDSS early-type green-valley galaxy sample lay on top of the colour tracks modeled with short $\tau_{\rm{q}}$, whereas the colours of late-type green-valley galaxies agree with larger $\tau_{\rm{q}}$ colour tracks. Therefore, a direct comparison between $\tau_{\rm{green}}$ and $\tau_{\rm{q}}$ cannot be made.

In the following section we investigate whether the green valley transition timescale can be explained by AGN feedback, morphological transformation or environment.


\section{Quenching mechanisms}


\subsection{Environment}

It has been found that galaxy morphology (e.g. \citealt{Poggianti99,Postman05,Bamford09}) and star formation rate (e.g. \citealt{Gomez03}, as well as the quenched galaxy fraction e.g. \citealt{Kauffmann03,Baldry06,Peng12}) correlate with environment, with star-forming disc galaxies preferentially residing in low-density regions, while quenched elliptical galaxies tend to reside in high-density regions. This suggests that environment may cause the galaxy to transition from the blue cloud to the red sequence. \citet{Schawinski14}, as well as \citet{Skibba09}, reported that late-type galaxies (analogous to disc-type in this work) in the green-valley reside almost exclusively in high-mass haloes ($\geq 10^{12}\Msun$), indicating that the quenching mechanism of these galaxies may be associated with environment, and that these galaxies are able to quench without changing morphology. In this section, we investigate this by analysing the relation between the galaxy's stellar mass, host halo mass and the green valley transition timescale, $\tau_{\rm{Green}}$, for both disc and elliptical RS satellite galaxies.

\begin{figure} 
\centering
\includegraphics[angle=0,width=0.46\textwidth]{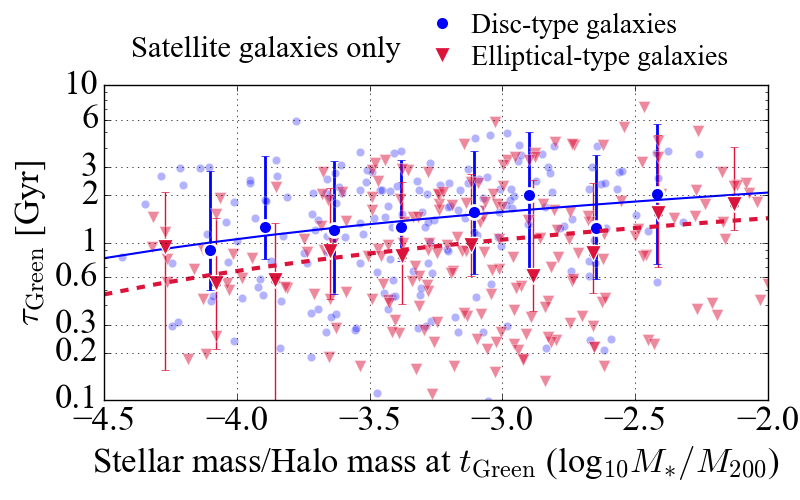}
\caption{Green valley transition timescale, $\tau_{\rm{Green}}$, of satellite galaxies as a function of the ratio between stellar mass and host halo mass, $\log_{10}M_{*}/M_{200}$, at the time when the galaxy left the blue cloud, $t_{\rm{Green}}$. The data points with errorbars show the median $\tau_{\rm{Green}}$ values in bins of $\log_{10}M_{*}/M_{200}$ for galaxies that have been satellites since $t_{\rm{Green}}$ and at $z{=}0$ have $\kappa_{\rm{co}}\ge 0.4$ (shown as blue dots) or $\kappa_{\rm{co}}< 0.4$ (shown as red dots). The error bars correspond to the $25$th$-75$th percentiles of the $\tau_{\rm{Green}}$ distribution in each mass bin. The lines correspond to the best-fit linear regressions. Red, satellite galaxies tend to have quenched more rapidly if they are elliptical and if their stellar mass is smaller relative to the total mass of their host halo. This relation is mostly driven by a dependence on halo mass, rather than on stellar mass.}
\label{sat_quenching}
\end{figure}

\begin{figure} 
\centering
\includegraphics[angle=0,width=0.48\textwidth]{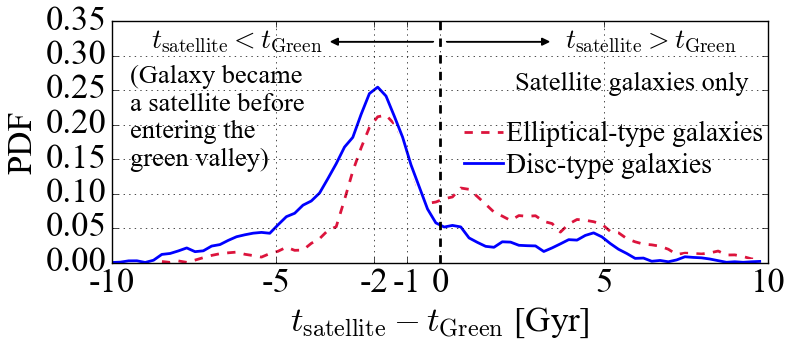}
\caption{PDF of the difference between the last time a red-sequence galaxy entered the green valley, $t_{\rm{Green}}$, and the first time it became a satellite ($t_{\rm{satellite}}$). The blue (red) lines correspond to disc (elliptical) galaxies. The times correspond to cosmic age (in Gyr), therefore negative time differences indicate that the first time a galaxy became a satellite occurred {\it{before}} the galaxy entered the green valley. Most red-sequence disc satellite galaxies became satellites ${\approx}2$ Gyr before entering the green valley. The PDF for elliptical satellite galaxies also shows a predominant peak at $t_{\rm{satellite}}-t_{\rm{Green}}{\approx}2$ Gyr, but roughly half the sample became satellites after entering the green valley.}
\label{sat_quenching_histogram_1}
\end{figure}

We select all $z=0$ RS satellite galaxies that, at the time the galaxy became green ($t_{\rm{Green}}$), were also satellites (corresponding to $70\%$ of the satellite sample), and calculate the ratio between the satellite galaxy's stellar mass and the host halo mass, $\log_{10}M_{*}/M_{200}$, at $t_{\rm{Green}}$. Throughout this section host halo mass refers to the halo mass\footnote{The halo mass, $M_{200}$, is defined to be total the mass contained within the virial radius, $R_{200}$, where the mean internal density is 200 times the critical density, $\rho_{\rm{crit}}$.} of the central galaxy within which the satellite galaxies are embedded. Fig.~\ref{sat_quenching} shows the green valley transition timescale, $\tau_{\rm{Green}}$, as a function of $\log_{10}M_{*}/M_{200}$ for RS satellite galaxies that are discs (blue dots) and ellipticals (red dots) at $z=0$. It can be seen that for both disc and elliptical satellite galaxies, $\tau_{\rm{Green}}$ tends to increase with $M_{*}/M_{200}$. This is highlighted by the best-fitting linear regression of the $\tau_{\rm{Green}}-\log_{10}M_{*}/M_{200}$ relation for elliptical (red dashed line) and disc (blue solid line) satellite galaxies. We find that $\tau_{\rm{Green}}$ correlates more strongly with $M_{200}$ than with $M_{*}$, so that satellite galaxies embedded in cluster-size haloes transition the green valley faster than their same-stellar mass satellite counterparts residing in lower mass haloes.

Fig.~\ref{sat_quenching} shows that satellite galaxies embedded in clusters ($\log_{10}M_{*}/M_{200}{<}-4$ with haloes more massive than $10^{14}\Msun$) tend to cross the green valley over shorter timescales than satellite galaxies residing in lower-mass haloes, regardless of the galaxy's morphological type. However, the spread in the $\tau_{\rm{Green}}$ distribution is large. In the case of central galaxies, although most of them reside in cluster-sized haloes, we do not find any correlation between $\tau_{\rm{Green}}$ and halo mass or $M_{*}/M_{200}$. This is in agreement with previous studies that suggest that environmental quenching is mostly associated with satellite galaxies rather than centrals (e.g., \citealt{Peng10,Darvish16}). Central galaxies show properties that are more associated with the galaxies' stellar mass (i.e. on average more massive galaxies are redder and less star-forming, e.g. \citealt{Kauffmann04,Baldry06}) and become quiescent independently of their host environment (\citealt{Peng12}). In this work however, we do not find a strong correlation between $\tau_{\rm{Green}}$ and stellar mass for centrals.

\begin{figure} 
\centering
\includegraphics[angle=0,width=0.48\textwidth]{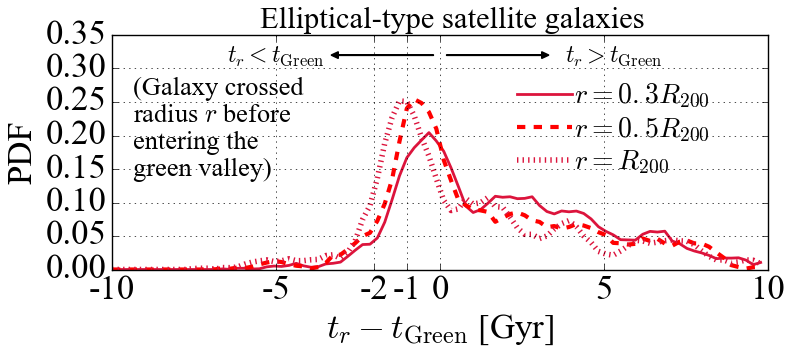}
\includegraphics[angle=0,width=0.48\textwidth]{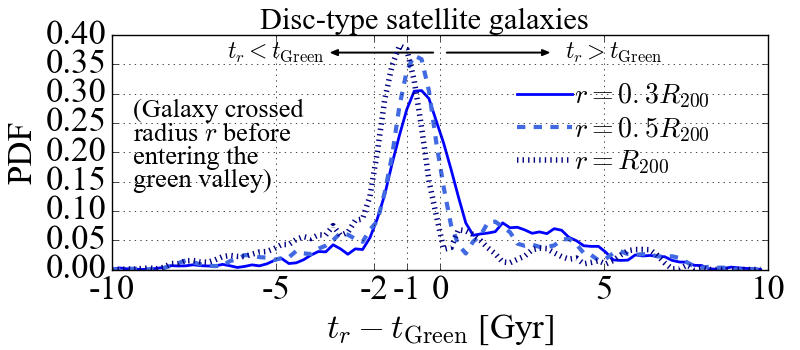}
\caption{PDF of the difference between the last time a red-sequence satellite galaxy entered the green valley, $t_{\rm{Green}}$, and the first time a galaxy crossed the radius $r$ (with $r=0.1R_{200}$, $r=0.5R_{200}$ or $r=R_{200}$, shown as solid, dashed and dotted lines, respectively). The top (bottom) panel corresponds to the time distribution of elliptical (disc) galaxies. Note that the times correspond to cosmic age, therefore negative time differences imply that the last time a galaxy crossed the radius $r$ occurred {\it{before}} the galaxy entered the green valley. The panels show that the closer a satellite is found to the central galaxy, the higher the probability that the galaxy has already entered the green valley. There is not a particular radius for which most galaxies redden {\it{at the same time}} they cross it. We find that most elliptical satellite galaxies had entered the green valley by the time they reached $0.5R_{200}$, whereas most disc galaxies had become green by the time they reached $0.3R_{200}$.}
\label{sat_quenching_histograms}
\end{figure}

We next investigate whether the time satellite galaxies left the blue cloud correlates with either the last time they became satellites, or the last time they reached a particular distance to their host halo's centre. To do so, we determine the {\it{first time}} a galaxy became a satellite ($t_{\rm{satellite}}$) and the {\it{first time}} ($t_{r}$) a galaxy reached a radius $r$, with $r$ being a fraction of the host halo's virial radius: $0.3\times R_{200}$, $0.5\times R_{200}$ or $R_{200}$. We compare these times with $t_{\rm{Green}}$, the last time the galaxy entered the green valley, and show the PDFs of $t_{\rm{satellite}}{-}t_{\rm{Green}}$ in Fig.~\ref{sat_quenching_histogram_1} and of $t_{r}{-}t_{\rm{Green}}$ in Fig.~\ref{sat_quenching_histograms}. Note that a galaxy becomes a satellite when it resides in a subhalo that is identified as a gravitationally self-bound aggregation of at least 50 particles (\citealt{Springel01}) within a Friends-of-Friends halo, so there is no direct link between the galaxy crossing $R_{200}$ and becoming a satellite (hence the difference between Figs. \ref{sat_quenching_histogram_1} and \ref{sat_quenching_histograms}). In the figures the times correspond to cosmic age, therefore $t_{\rm{satellite}}{<}t_{\rm{Green}}$ (negative time differences) indicate that the first time a galaxy became a satellite occurred {\it{before}} the galaxy entered the green valley. Similarly $t_{\rm{r}}{<}t_{\rm{Green}}$ means that the first time a galaxy reached the radius $r$ occurred {\it{before}} the galaxy entered the green valley.

Fig.~\ref{sat_quenching_histogram_1} shows that most RS disc satellite galaxies became satellites ${\approx}2$ Gyr before entering the green valley, with $80\%$ of them having $t_{\rm{satellite}}{<}t_{\rm{Green}}$. For elliptical satellite galaxies, roughly half the sample became satellites before entering the green valley. 

The top and bottom panels of Fig.~\ref{sat_quenching_histograms} show that the smaller the halocentric radius that satellite galaxies are crossing for the first time, the higher the probability that they have already entered the green valley (before reaching $r$). This can be seen by comparing the $t_{r}-t_{\rm{Green}}$ distributions for different values of $r$. In the top panel, the fraction of elliptical satellite galaxies that crossed $r=R_{200}$ {\it{before}} entering the green valley is $50\%$, decreasing to $30\%$ for $r=0.3R_{200}$. The bottom panel shows that $80\%$ of disc satellite galaxies crossed the radius $r=R_{200}$ {\it{before}} entering the green valley, decreasing to $56\%$ for $r=0.3R_{200}$. This indicates that the closer a satellite is found to the central galaxy, the higher the probability that the galaxy is no longer blue, and the probability is higher for elliptical rather than disc satellite galaxies.

\subsection{AGN feedback}

\begin{figure} 
\centering
\includegraphics[angle=0,width=0.48\textwidth]{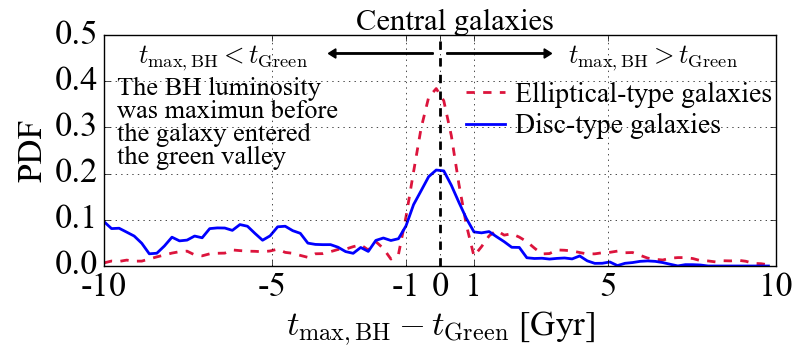}
\includegraphics[angle=0,width=0.48\textwidth]{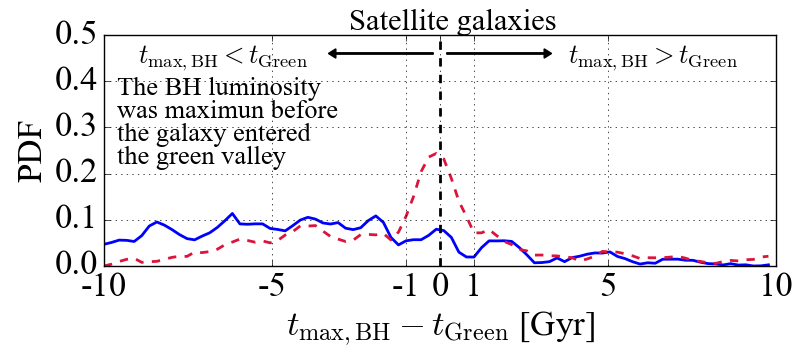}
\caption{PDF of the difference between the last time a red-sequence galaxy entered the green valley, $t_{\rm{Green}}$, and the time the galaxy's time-averaged central black hole luminosity was maximum ($t_{\rm{max,BH}}$). The blue (red) lines correspond to the time distribution of disc (elliptical) galaxies. The top panel shows the distribution for central galaxies and the bottom panel for satellites. For elliptical red-sequence galaxies, AGN feedback and black hole growth are correlated, particularly for centrals. For disc centrals the correlation is weaker and undetectable for satellites.}
\label{BH_quenching}
\end{figure}

The large fraction of AGN host galaxies that are found in the green valley (e.g. \citealt{Hickox09,Schawinski10}) can be viewed as an indication that there is a link between AGN activity and the process of quenching that moves a galaxy from the blue cloud to the red sequence. In this section we investigate this and analyze whether there is a correlation between the last time the galaxy entered the green valley, $t_{\rm{Green}}$, and the time the time-averaged central black hole (BH) luminosity is maximum, $t_{\rm{max,BH}}$. We calculate the black hole luminosities from the black hole accretion rates via the relation $L_{\rm{BH}}/L_{\odot}=1.65\times 10^{12}\dot{M}_{\rm{BH}}/(\Msun/\rm{yr})$ (where we assumed a radiative efficiency of $10\%$; \citealt{Shakura73}). For $\dot{M}_{\rm{BH}}$ we use the time-average black hole mass accretion rate ($\dot{M}_{\rm{BH}}=\Delta M_{\rm{BH}}/\Delta t$) between two consecutive snipshots. To avoid large time-averaged $\dot{M}_{\rm{BH}}$ driven by BH-BH mergers, we discard those where the $\Delta{M}_{\rm{BH}}$ is equal to the largest mass of a neighboring BH within 50 kpc in the previous snipshot. Since most massive BHs about to merge are within 5 kpc of each other in the previous snipshot, $t_{\rm{max,BH}}$ does not change with neighboring BH distance lower than 50 kpc.

The top and bottom panels of Fig.~\ref{BH_quenching} show the PDFs of the difference between $t_{\rm{Green}}$ and $t_{\rm{max,BH}}$ for disc (blue lines) and elliptical (red lines) RS galaxies that are centrals and satellites at $z=0$, respectively. The distribution of elliptical galaxies peaks at $t_{\rm{max,BH}}-t_{\rm{Green}}{\approx}0$, indicating that many galaxies became green when the time-averaged BH luminosity was highest. We find that for $48\%$ ($34\%$) of the elliptical central (satellite) galaxy population the BH luminosity was maximum {\it{at the same time}} the galaxies entered the green valley (i.e. $|t_{\rm{max,BH}}-t_{\rm{Green}}|{<}1$ Gyr). In the case of discs, this occurs for $30\%$ ($10\%$) of the central (satellite) population. We conclude that the AGN feedback contributed to the quenching of central galaxies, particularly for ellipticals, and of elliptical satellites.


\subsection{Morphological transformation}\label{color_morpho_evolution_sec}

\begin{figure} 
\begin{center}
\includegraphics[angle=0,width=0.48\textwidth]{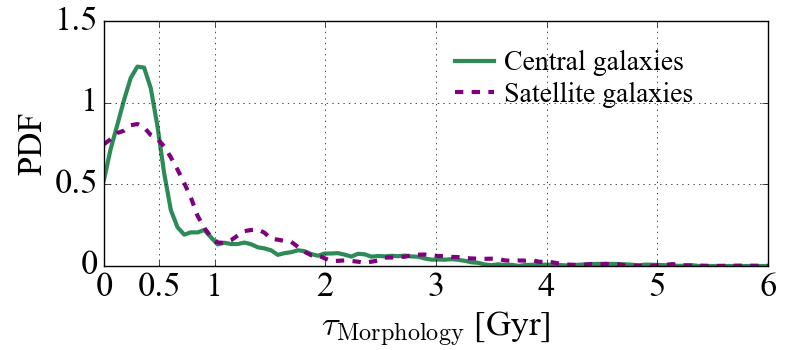}
\end{center}
\vspace{-0.4cm}
\caption{PDF of the time, $\tau_{\rm{Morphology}}$, red-sequence galaxies (both centrals and satellites) take to change morphology from disc to elliptical. The solid line corresponds to central galaxies and the dashed line to satellites. The figure shows a clear population of central and satellite galaxies that change morphology in less than 1 Gyr.}
\label{histogram_morpho_change}
\end{figure}

\begin{figure} 
\begin{center}
\includegraphics[angle=0,width=0.48\textwidth]{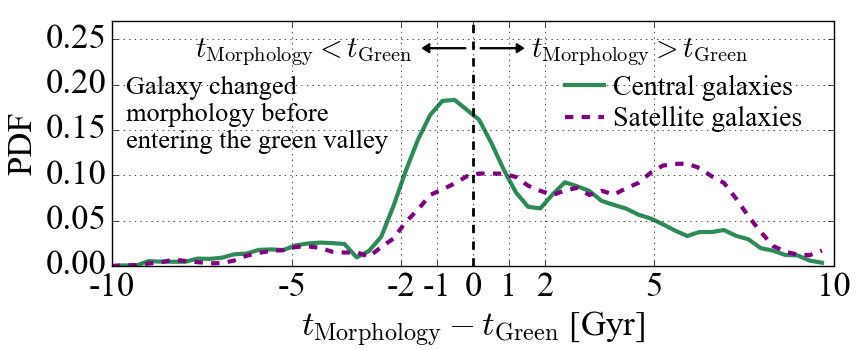}
\end{center}
\vspace{-0.4cm}
\caption{PDF of the difference in time between the last time red-sequence (RS) galaxies changed morphology (had $\kappa_{\rm{co}}>0.4$, $t_{\rm{Morphology}}$) and the last time RS galaxies became green ($t_{\rm{Green}}$). The times correspond to cosmic age. If $t_{\rm{Morphology}}<t_{\rm{Green}}$, the RS galaxies changed morphology before becoming green. Solid lines correspond to the time distribution of central galaxies whereas the dashed line corresponds to satellites. A large fraction ($45\%$) of RS galaxies that are centrals tend to have changed morphology before entering the green valley, whereas most satellites ($75\%$) tend to become green before changing morphology.}
\label{times_diff}
\end{figure}

Previous works have found that the formation of a bulge in a disc galaxy precedes any significant reduction of star formation, implying that morphological transformation occurs before quenching (e.g. \citealt{Fang13,Bremer18}). In this section we analyse the morphological history of RS EAGLE galaxies, and investigate the temporal correspondence between changes in morphology and colour. We select RS galaxies that are elliptical at $z=0$ but disc at an earlier time, corresponding to a subsample of 583 galaxies ($86\%$ of the $z=0$ RS elliptical galaxy sample). There are 92 RS elliptical galaxies that did not change morphology and were elliptical between $z=0$ and 4. 

We calculate $\tau_{\rm{Morphology}}$, the time galaxies took to change from disc to elliptical. $\tau_{\rm{Morphology}}$ is defined as the difference between the last time galaxies had a corotational parameter, $\kappa_{\rm{co}}{>}0.5$, and the last time galaxies had $\kappa_{\rm{co}}{>}0.3$. The limits $\kappa_{\rm{co,max}}{=}0.5$ and $\kappa_{\rm{co,min}}{=}0.3$ are ad-hoc, but we find that if we change them to $\kappa_{\rm{co,max}}{=}0.4$ and $\kappa_{\rm{co,min}}{=}0.3$ we obtain the same qualitative results. We find that $72\%$ of the subsample (422 galaxies) reaches values of $\kappa_{\rm{co}}{>}0.5$ and $\kappa_{\rm{co}}{<}0.3$, so we restrict our analysis of $\tau_{\rm{Morphology}}$ to this subsample that undergoes a clearer morphological transformation.

Fig.~\ref{histogram_morpho_change} shows the distribution of $\tau_{\rm{Morphology}}$ for central (solid line) and satellite (dashed line) RS galaxies. It can be seen that the morphological transformation from disc to elliptical predominantly occurs in less than 1 Gyr. Central and satellite galaxies change morphology on median timescales of ${\approx}300$ Myr and ${\approx}420$ Myr, respectively.

We compare $t_{\rm{Green}}$ with the time RS galaxies changed morphology, $t_{\rm{Morphology}}$, defined as the last time $\kappa_{\rm{co}}{>}0.4$. The $t_{\rm{Morphology}}{-}t_{\rm{Green}}$ distribution is shown in Fig.~\ref{times_diff}, where $t_{\rm{Morphology}}{<}t_{\rm{Green}}$ (negative time differences) indicates that galaxies changed morphology before becoming green. Similarly, Fig.~\ref{times_diff} means that galaxies entered the green valley before changing morphology. The figure shows that a large fraction ($45\%$) of central galaxies tend to change morphology {\it{before}} entering the green valley. But this is not the case for satellites, which preferentially became green before changing morphology.

\begin{figure} 
\begin{center}
\includegraphics[angle=0,width=0.48\textwidth]{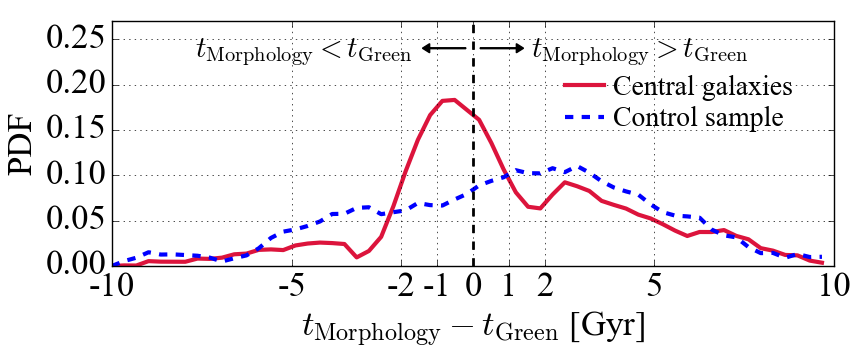}\\
\vspace{-0.7cm}
\includegraphics[angle=0,width=0.48\textwidth]{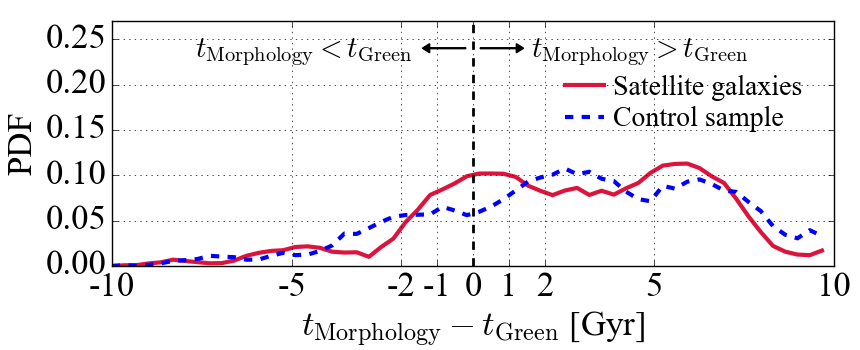}
\vspace{-0.5cm}
\caption{Same as Fig. 12. Top panel: comparison between the PDF of $t_{\rm{Morphology}}-t_{\rm{Green}}$ of red-sequence (RS) central galaxies (red solid line) and a control sample (blue dashed line). Bottom panel: same as top panel but for satellite galaxies.}
\label{color_morpho_analysis}
\end{center}
\end{figure}

Fig.~\ref{times_diff} also shows a prominent peak in the PDF of RS centrals at $t_{\rm{Morphology}}{-}t_{\rm{Green}}{\approx}-0.5$ Gyr, which indicates that there is a non-negligible number of central galaxies where the morphological transformation occurs {\it{at approximately the same time}} as the reddening. We find that for $36\%$ ($17\%$) of the central (satellite) galaxy subsample $|t_{\rm{Morphology}}{-}t_{\rm{Green}}|{<}-0.5\pm 1$ Gyr, from which we could conclude that reddening and morphological transformation are physically connected. However, it is possible that these fractions do not reflect a physical connection, but merely result from how the galaxy subsample is built. To test this we compare the PDFs of $t_{\rm{Morphology}}-t_{\rm{Green}}$ of RS galaxies to those of a control sample. The control sample is selected as follows. We loop over the RS galaxies of the subsample and for each galaxy we randomly select two galaxies $i$ and $j$ from the same subsample. We determine the difference between the time $t_{\rm{Green}}$ of galaxy $i$ and the time $t_{\rm{Morphology}}$ of galaxy $j$, and calculate the PDF of $t_{\rm{Morphology},j}{-}t_{\rm{Green},i}$. Any correlation between $t_{\rm{Morphology},j}$ and $t_{\rm{Green},i}$ is coincidental due to the selection criteria of the subsample and not indicative of a physical relation, because the morphological transformation and reddening occur in different galaxies.

Fig.~\ref{color_morpho_analysis} shows a comparison between the $t_{\rm{Morphology}}{-}t_{\rm{Green}}$ PDFs of central galaxies and the control sample (top panel) and of satellite galaxies and their control sample (bottom panel). Note that for each case, central or satellite galaxies, the control sample is built so that it contains only central galaxies or only satellites, respectively. The top panel shows that the PDF of the control sample does not exhibit a peak at $t_{\rm{Morphology}}{-}t_{\rm{Green}}{\approx}-0.5$ Gyr. A Kolmogorov-Smirnov (KS) test\footnote{The Kolmogorov-Smirnov test returns a $p$-value that indicates the probability that two distributions represent the same sample.} indicates that the distributions differ at the $p=0.03$ level, confirming that the enhanced fraction of central galaxies for which the morphological transformation occurs close in time to the reddening is indicative of a physical connection. 

As noted above, for $36\%$ of the central galaxies $|t_{\rm{Morphology}}{-}t_{\rm{Green}}|{<}-0.5\pm 1$ Gyr. However, this fraction likely includes galaxies for which the correlation of $t_{\rm{Morphology}}$ and $t_{\rm{Green}}$ is coincidental. To correct for that we determine the excess of central galaxies with respect to the control sample, and find it to be $18\%$. 

From the bottom panel of Fig.~\ref{color_morpho_analysis} it can be seen that there is not a large difference between the $t_{\rm{Morphology}}{-}t_{\rm{Green}}$ PDFs of satellites and the control. Indeed, a KS test yields a probability $p=0.34$ that the two samples are drawn from the same distribution.


\section{Conclusions}

The morphology of a galaxy is a fundamental property that can be used to infer possible evolutionary paths. In this work we have investigated the origin of the red sequence (RS) galaxy population in the EAGLE cosmological hydrodynamical simulation. We followed the evolution in colour and morphology of the progenitors of RS galaxies more massive than $10^{10}\Msun$. We quantified galaxy morphology using a kinematic parameter, $\kappa_{\rm{co}}$, that tracks the rotational support of the stellar component. We showed that $\kappa_{\rm{co}}$ separates galaxies in the angular momentum-mass plane along lines that observationally have been shown to have different visual morphology (Fig.~\ref{comparison}), with galaxies of large $\kappa_{\rm{co}}$ values being mostly late-types with a disc-component, and galaxies of low $\kappa_{\rm{co}}$ values being mostly ellipticals.

We used the $u^{*}-r^{*}$ intrinsic colours of \citet{Trayford16} to separate the RS galaxy population from the blue cloud. By fitting a double gaussian to the $u^{*}-r^{*}$ colour distribution (Fig.~\ref{fig_images}), we defined colour boundaries as $1\sigma$ away from the population mean colour as a function of redshift. Most central, red and green galaxies are elliptical-type, while central blue galaxies are predominately disc-type. For the case of satellites, red and green galaxies are morphologically diverse (Fig.~\ref{kappa_fraction}). 

The evolution in colour and morphology of the progenitors of RS galaxies indicates that the time galaxies move onto the RS depends on their morphology (Fig.~\ref{morpho_distribution}). Disc-type galaxies mostly joined the RS during the last 3 Gyr, whereas elliptical-types populated the RS earlier, 5 Gyr ago (at which point many were still disc-type, Fig.~\ref{fE_fR_evolution}). We calculated the time galaxies took to transition the green valley, and found that disc-type galaxies move from the blue cloud to the RS over a median timescale of ${\approx}1.5$ Gyr, whereas elliptical-type galaxies quenched slightly faster, over a timescale of ${\approx}1$ Gyr (Fig.~\ref{quench_time}). 

We investigated whether environment, AGN feedback and morphological transformation are mechanisms responsible for the transition from the blue cloud to the RS. The green-valley transition timescale of RS galaxies that are satellites correlates with the ratio between stellar mass and host halo mass at the time when the galaxy entered the green valley (Fig.~\ref{sat_quenching}), indicating that satellite galaxies embedded in cluster-sized haloes cross the green valley faster (in less than 1 Gyr) than same-mass satellites in lower-mass haloes. Most satellite RS galaxies became satellites before entering the green valley (Fig.~\ref{sat_quenching_histogram_1}). Thus, the quenching of satellite RS galaxies is partly caused by satellite-specific environmental processes. Most elliptical-type satellite galaxies entered the green valley by the time they reached $0.5R_{200}$, whereas most disc-types became green by the time they reached $0.3R_{200}$ (Fig.~\ref{sat_quenching_histograms}).

The time the central black hole luminosity was maximum in central RS galaxies, $t_{\rm{max,BH}}$, correlates with the time the galaxies became green, $t_{\rm{Green}}$. This is particularly the case for central elliptical-types, with roughly half of them ($30\%$ for satellite elliptical-types) having $|t_{\rm{max,BH}}-t_{\rm{Green}}|<1$ Gyr. For disc-types centrals the correlation is weaker and it is undetectable for satellites (Fig.~\ref{BH_quenching}). Hence, for elliptical-type, but not for disc-type RS galaxies, AGN feedback appears to contribute significantly to the quenching process.

The morphological transformation from disc- to elliptical-type of the progenitors of RS galaxies is rapid for central and satellite galaxies, where most of the sample changed morphology in less than 1 Gyr (Fig.~\ref{histogram_morpho_change}). Also, while a large fraction ($45\%$) of central RS galaxies changed morphology before entering the green valley, most satellites became green before changing morphology  (Fig.~\ref{times_diff}). For the progenitors of central RS galaxies, reddening and morphology transformation are significantly correlated in time (Fig.~\ref{color_morpho_analysis}), but this is not the case for satellites. However, the fraction of central RS galaxies that underwent a morphological transformation as they entered the green valley is $18\%$ and negligible for satellites. Thus, there is a weak connection between the mechanisms responsible for the colour and morphology change of the progenitors of RS ellipticals, but only for centrals.




\section*{Acknowledgments}

We thank the anonymous referee for fruitful comments that substantially improved the original manuscript. We are grateful to the EAGLE team for putting together a great set of simulations, and sharing the merger tree catalogs. This work used the DiRAC Data Centric system at Durham University, operated by the Institute for Computational Cosmology on behalf of the STFC DiRAC HPC Facility (www.dirac.ac.uk). This equipment was funded by BIS National E-infrastructure capital grant ST/K00042X/1, STFC capital grant ST/H008519/1, and STFC DiRAC Operations grant ST/K003267/1 and Durham University. DiRAC is part of the National E-Infrastructure. The EAGLE simulations were performed using the DiRAC-2 facility at Durham, managed by the ICC, and the PRACE facility Curie based in France at TGCC, CEA, Bruyeres-le-Chatel. This work was supported by the Netherlands Organisation for Scientific Research (NWO) through VICI grant 639.043.409. We have benefited greatly from the public available programming language {\tt{PYTHON}}, including the {\tt{NUMPY}}, {\tt{MATPLOTLIB}}, {\tt{SCIPY}}, {\tt{H5PY}} and {\tt{astropy}} packages.

\bibliography{biblio}
\bibliographystyle{mn2e}

\end{document}